\documentclass[conference]{IEEEtran}

\usepackage{cite}
\usepackage[dvips]{graphicx}
\usepackage[cmex10]{amsmath}
\usepackage{algorithmic}
\usepackage{algorithm}
\usepackage{amssymb}
\usepackage{balance}

\DeclareMathOperator*{\argmin}{arg\,min}
\DeclareMathOperator*{\x}{\mathbf{x}}
\DeclareMathOperator*{\z}{\mathbf{z}}
\DeclareMathOperator*{\A}{\mathbf{A}}

\DeclareMathOperator*{\blambda}{\boldsymbol \lambda}
\DeclareMathOperator*{\R}{\mathbb{R}}
\DeclareMathOperator*{\so}{\mathbf{o}}
\DeclareMathOperator*{\m}{\mathbf{m}}
\DeclareMathOperator*{\J}{\mathbf{J}}

\begin{document}
\title{Multi-Block ADMM for \\Big Data Optimization in Smart Grid}

\author{\IEEEauthorblockN{Lanchao Liu and Zhu Han}
\IEEEauthorblockA{Department of Electrical and Computer Engineering\\
University of Houston,
Houston, TX, 77004}}


\maketitle

\begin{abstract}
In this paper, we review the parallel and distributed optimization algorithms based on alternating direction method of multipliers (ADMM) for solving ``big data'' optimization problem in smart grid communication networks. We first introduce the canonical formulation of the large-scale optimization problem. Next, we describe the general form of ADMM and then focus on several direct extensions and sophisticated modifications of ADMM from $2$-block to $N$-block settings to deal with the optimization problem. The iterative schemes and convergence properties of each extension/modification are given, and the implementation on large-scale computing facilities is also illustrated. Finally, we numerate several applications in power system for distributed robust state estimation, network energy management and security constrained optimal power flow problem.
\end{abstract}

\section{Introduction}
The unprecedented ``big data'', reinforced by communication and information technologies, presents us opportunities and challenges. On one hand, the inferential power of algorithms, which have been shown to be successful on modest-sized data sets, may be amplified by the massive dataset. Those data analytic methods for the unprecedented volumes of data promises to personalized business model design, intelligent social network analysis, smart city development, efficient healthcare and medical data management, and the smart grid evolution. On the other hand, the sheer volume of data makes it unpractical to collect, store and process the dataset in a centralized fashion. Moreover, the massive datasets are noisy, incomplete, heterogeneous, structured, prone to outliers, and vulnerable to cyber-attacks. The error rates, which are part and parcel of any inferential algorithm, may also be amplified by the massive data. Finally, the ``big data" problems often come with time constraints, where a medium-quality answer that is obtained quickly can be more useful than a high-quality answer that is obtained slowly. Overall, we are facing a problem in which the classic resources of computation such as time, space, and energy, are intertwined in complex ways with the massive data resources.

With the era of ``big data'' comes the need of parallel and distributed algorithms for the large-scale inference and optimization. Numerous problems in statistical and machine learning, compressed sensing, social network analysis, and computational biology formulates optimization problems with millions or billions of variables. Since classical optimization algorithms are not designed to scale to problems of this size, novel optimization algorithms are emerging to deal with problems in the ``big data'' setting. An incomprehensive list of such kind of algorithms includes block coordinate descent method \cite{PS09,YS09,Y12}\footnote{\cite{Y12} proposes a stochastic block coordinate descent method.} , stochastic gradient descent method \cite{LO08, MMAL10, FBCS11}, dual coordinate ascent method \cite{CKCSS08, ST13}, alternating direction method of multipliers (ADMM) \cite{BT97, BPCPE10} and Frank-Wolf method (also known as the conditional gradient method) \cite{RP14,SMMP13}. Each type of the algorithm on the list has its own strength and weakness. The list is sill growing and due to our limited knowledge and the fast develop nature of this active field of research, many efficient algorithms are not mentioned here.

In this paper, we focus on the application of ADMM for the ``big data'' optimization problem in smart grid communication networks. In particular, we consider the parallel and distributed optimization algorithms based on ADMM for the following convex optimization problem with a canonical form as
\begin{align}
\label{eqn:intro}
\min_{\x_1,\x_2, \ldots, \x_N}\quad & f(\x) = f_i({\x}_i) + \ldots + f_i({\x}_N),\nonumber \\
\text{s.t.}\quad &{\A}_i{\x}_i + \ldots + {\A}_N{\x}_N  = \mathbf{c}, \nonumber \\
&{\x}_i \in \mathcal{X}_i, \quad i = 1,\ldots,N,
\end{align}
where $\x = ({\x}_1^{\top}, \ldots, {\x}_N^{\top})^{\top}$, $\mathcal{X}_i \subset {\R}^{n_i} (i = 1,2, \ldots, N)$ are closed convex set, ${\A}_i \in {\R}^{m \times n_i}(i = 1,2, \ldots, N)$ are given matrices, $\mathbf{c} \in {\R}^{m}$ is a given vector, and $f_i: {\R}^{n_i} \rightarrow {\R}$ $(i = 1,2, \ldots, N)$ are closed convex proper but not necessarily smooth functions, where the non-smoothness functions are usually employed to enforce structure in the solution. Problem (\ref{eqn:intro}) can be extended to handle linear inequalities by introducing slack variables. Problem (\ref{eqn:intro}) finds wide applications in smart grid on distributed robust state estimation, network energy management and security constrained optimal power flow problem, which we will illustrated later.

Though many algorithms can be applied to deal with problem (\ref{eqn:intro}), we restrict our attention to the class of algorithms based on ADMM. The rest of this paper is organized as follows. Sec. \ref{sec:background} introduces the background of the ADMM and its two direct extensions for problem (\ref{eqn:intro}) to $N$ blocks. The limitations of those direct extensions are also addressed. Sec. \ref{sec:multiblock} gives three approaches based on Variable Splitting, ADMM with Gaussian back substitution and proximal Jacobian ADMM to the multi-block settings, respectively, for problem (\ref{eqn:intro}) with provable convergence. The applications of problem (\ref{eqn:intro}) in smart grid communication networks are described in Sec. \ref{sec:application}. Sec. \ref{sec:conclusion} summarized this paper.

\section{ADMM Background}
\label{sec:background}

In this section, we first introduce the general form of ADMM for optimization problem analogous to (\ref{eqn:intro}) with only two blocks of functions and variables. After that, we described two direct extensions of ADMM to multi-block setting.

\subsection{ADMM}
The ADMM was proposed in \cite{RA75}, \cite{DB76} and recently revisited by \cite{BPCPE10}. The general form of ADMM is expressed as
\begin{equation}
\label{eqn:StdAMDD}
\min_{\x_1 \in \mathcal{X}_1,\x_2 \in \mathcal{X}_2} f_1({\x}_1) + f_2({\x}_2) \quad \text{s.t.} \quad {\A}_1{\x}_1 + {\A}_2{\x}_2 = \mathbf{c}.
\end{equation}
The augmented Lagrangian for (\ref{eqn:StdAMDD}) is
\begin{align}
\mathcal{L}_{\rho} ({\x}_1,{\x}_2,{\blambda})  &= f_1({\x}_1) + f_2({\x}_2) - {\blambda}^{\top}( {\A}_1{\x}_1 + {\A}_2{\x}_2 - \mathbf{c}) \nonumber \\
&+ \frac{\rho}{2}\Vert {\A}_1{\x}_1 + {\A}_2{\x}_2 - \mathbf{c} \Vert_2^2,
\end{align}
where $\blambda \in {\R}^{m}$ is the Lagrangian multiplier and $\rho > 0$ is the parameter for the quadratic penalty of the constraints. The iterative scheme of ADMM embeds a Gauss-Seidel decomposition into iterations of ${\x}_1$ and ${\x}_2$ as follows
\begin{equation}
\label{eqn:update}
\left\{ \begin{array}{l}
{\x}_1^{k+1} = \argmin_{{\x}_1}\mathcal{L}_{\rho} ({\x}_1,{\x}_2^{k},{\blambda}^{k}),\\
{\x}_2^{k+1} = \argmin_{{\x}_2}\mathcal{L}_{\rho} ({\x}_1^{k+1},{\x}_2,{\blambda}^{k}),\\
{\blambda}^{k+1} = {\blambda}^{k} - {\rho}({\A}_1{\x}_1^{k+1} + {\A}_2{\x}_2^{k+1} - \mathbf{c}).
\end{array} \right.
\end{equation}
where in each iteration, the augmented Lagrangian is minimized over ${\x}_1$ and ${\x}_2$ separately. In (\ref{eqn:update}), functions $f_1$ and $f_2$ as well as variables ${\x}_1$ and ${\x}_2$ are treated individually, so easier subproblems can be generated. This feature is quite attractive and advantageous for a broad spectrum of applications. The convergence of ADMM for convex optimization problem with two blocks of variables and functions has been proved in \cite{BT97}, \cite{BPCPE10},  and the iterative scheme is illustrated in Algorithm \ref{alg:A1}. Algorithm \ref{alg:A1} can deal with multi-block case when auxiliary variables are introduced, which will be described in Sec. \ref{sec:VSADMM}.

\begin{algorithm}[t]
\caption{Two-block ADMM}\label{alg:A1}
\begin{algorithmic}
\STATE Initialize: ${\x}^{0}$, ${\blambda}^{0}$, $\rho>0$;

\FOR{$k=0,1,\ldots$}

\STATE ${\x}_1^{k+1} = \argmin_{{\x}_1}\mathcal{L}_{\rho} ({\x}_1,{\x}_2^{k},{\blambda}^{k})$;

\STATE ${\x}_2^{k+1} = \argmin_{{\x}_2}\mathcal{L}_{\rho} ({\x}_1^{k+1},{\x}_2,{\blambda}^{k})$;

\STATE ${\blambda}^{k+1} = {\blambda}^{k} - {\rho}({\A}_1{\x}_1^{k+1} + {\A}_2{\x}_2^{k+1} - \mathbf{c})$;

\ENDFOR

%\STATE Output ${\x}$, ${\blambda}$;
\end{algorithmic}
\end{algorithm}

\subsection{Direct Extensions to Multi-block Setting}
The ADMM promises to solve the optimization problem (\ref{eqn:intro}) with the same philosophy as algorithm \ref{alg:A1}. In the following, we present two kinds of direct extensions, Gauss-Seidel and Jacobian, for multi-block ADMM. To be specific, we first give the augmented Lagrangian function of problem (\ref{eqn:intro})
\begin{align}
\label{eqn:ALM}
\mathcal{L}_{\rho} ({\x}_1,\ldots,{\x}_N,{\blambda})  = &\sum_{i=1}^{N}f_i({\x}_i) - {\blambda}^{\top}( \sum_{i=1}^{N} {\A}_i{\x}_i - \mathbf{c})\\ \nonumber
&+ \frac{\rho}{2}\Vert \sum_{i=1}^{N} {\A}_i{\x}_i - \mathbf{c} \Vert_2^2.
\end{align}

\subsubsection{Gauss-Seidel}
Intuitively, a natural extension of the classical Gauss-Seidel setting ADMM from $2$ blocks to $N$ blocks is a straightforward replacement of the two-block alternating minimization scheme by a sweep of update of ${\x}_i$ for $i = 1,2,\ldots,N$ sequentially. In particular, at iteration $k$, the update scheme for ${\x}_i$ is
\begin{equation}
{\x}_i = \argmin_{{\x}_i}\mathcal{L}_{\rho}(\{{\x}_j^{k+1}\}_{j<i},{\x}_i,\{{\x}_j^{k}\}_{j>i},{\blambda}^k),
\end{equation}
where $\{{\x}_j\}_{j<i}$ denotes the set of variables prior to $i$. The augmented Lagrangian function (\ref{eqn:intro}) is split and updated alternatingly. The direct Gauss-Seidel type extension can be illustrated in Algorithm \ref{alg:A2}.

\textbf{Remark:} Algorithm \ref{alg:A2} has been utilized in practical problems \cite{YAJWY12, MX11, HCB14} despite a lack of rigourous proof for the convergence. Actually, the convergence of Gauss-Seidel multi-block ADMM is not well understood and is ambiguous for a long time: Neither affirmative convergence proof nor counter examples for convergence failure are shown in the literature. Recently, \cite{CBYX13} has shown that the direct extension of Gauss-Seidel mulit-block ADMM is not necessarily convergent. \cite{MZ12} prove the convergence of Algorithm \ref{alg:A2} with sufficient small step size for Lagrangian multiplier update and additional assumptions on the problem (\ref{eqn:intro}). \cite{CBYX14} conjectures that an independent uniform random permutation of the update order for blocks in each iteration will result in a convergent iteration scheme.  \cite{BXM12,MTXMSZ14} proposed some slightly modified version of Algorithm \ref{alg:A2} with provable convergence and competitive iteration simplicity and computing efficiently, which we will illustrate later in Sec. \ref{sec:GSADMM}.

\begin{algorithm}[t]
\caption{Gauss-Seidel Multi-block ADMM}\label{alg:A2}
\begin{algorithmic}
\STATE Initialize: ${\x}^{0}$, ${\blambda}^{0}$, $\rho>0$;

\FOR{$k=0,1,\ldots$}

\FOR{$i=1,\ldots,N$}

\STATE \COMMENT{${\x}_i$ is updated \textbf{sequentially}.}

\STATE ${\x}_i^{k+1} = \argmin_{{\x}_i}\mathcal{L}_{\rho}(\{{\x}_j^{k+1}\}_{j<i},{\x}_i,\{{\x}_j^{k}\}_{j>i},{\blambda}^k)$;

\ENDFOR

\STATE ${\blambda}^{k+1} = {\blambda}^{k} - {\rho}(\sum_{i=1}^{N} {\A}_i{\x}_i^{k+1} - \mathbf{c})$;

\ENDFOR

%\STATE Output ${\x}$, ${\blambda}$;
\end{algorithmic}
\end{algorithm}

\subsubsection{Jacobian}
Another possible iterative scheme for the $N$ blocks ADMM is the Jacobian type update, which performs the update of ${\x}_i$ in a parallel coordinate fashion for $i = 1, \ldots, N$. In particular, the update of ${\x}_i$ is calculated as:
\begin{equation}
{\x}_i = \argmin_{{\x}_i}\mathcal{L}_{\rho}({\x}_i,\{{\x}_j^{k}\}_{j \neq i},{\blambda}^k),
\end{equation}
where $\{{\x}_j^{k}\}_{j \neq i}$ denotes the set of variables except for ${\x}_i$. Different from the iterative scheme of Algorithm \ref{alg:A2} that the update of ${\x}_i$ has to be performed sequentially one after another, the iterations in the Jacobian ADMM can be performed concurrently, i.e. all ${\x}_i$ can be updated in a parallel fashion. This advantage makes the Jacobian type ADMM preferred for parallel implementation, and the direct Jacobian type extension can be illustrated in Algorithm \ref{alg:A3}.

\textbf{Remark} Though Algorithm \ref{alg:A3} is more computational efficient in the sense of parallelization, \cite{BLX13} shows that Algorithm \ref{alg:A3} is not necessarily convergent in the general case, even in the 2 blocks case. \cite{WMZW14} proves that if matrices ${\A}_i$ are mutually near-orthogonal and have full column-rank, the Algorithm \ref{alg:A3} converges globally. A proximal Jacobian ADMM is also proposed in \cite{WMZW14} with provable convergence, which we will illustrate later in Sec. \ref{sec:PJADMM}

\begin{algorithm}[t]
\caption{Jacobian Multi-block ADMM}\label{alg:A3}
\begin{algorithmic}
\STATE Initialize: ${\x}^{0}$, ${\blambda}^{0}$, $\rho>0$;

\FOR{$k=0,1,\ldots$}

\FOR{$i=1,\ldots,N$}

\STATE \COMMENT{${\x}_i$ is updated \textbf{concurrently}.}

\STATE ${\x}_i^{k+1} = \argmin_{{\x}_i}\mathcal{L}_{\rho}({\x}_i,\{{\x}_j^{k}\}_{j \neq i},{\blambda}^k)$;

\ENDFOR

\STATE ${\blambda}^{k+1} = {\blambda}^{k} - {\rho}(\sum_{i=1}^{N} {\A}_i{\x}_i^{k+1} - \mathbf{c})$;

\ENDFOR

%\STATE Output ${\x}$, ${\blambda}$;
\end{algorithmic}
\end{algorithm}

\section{Multi-block ADMM}
\label{sec:multiblock}
In this section, we introduce several sophisticated modifications of ADMM, Variable splitting ADMM \cite{BT97, BPCPE10, MJM10}, ADMM with Gaussian Back Substitution \cite{BXM12,BMX12} and Proximal Jacobian ADMM \cite{WMZW14,BMX13}, to deal with the multi-block setting.
\subsection{Variable Splitting ADMM}
\label{sec:VSADMM}
To solve the optimization problem (\ref{eqn:intro}), we can apply the variable splitting \cite{BT97, BPCPE10, MJM10} to deal with the multi-block variables. In particular, the optimization problem (\ref{eqn:intro}) can be reformulated by introducing auxiliary variable $\z$
\begin{align}
\label{eqn:VS}
\min_{\x,\z}\quad & \sum_{i=1}^{N} f_i({\x}_i) + I_{\mathcal{Z}}({\z}),\nonumber \\
\text{s.t.}\quad &{\A}_i{\x}_i + {\z}_i  = \frac{\mathbf{c}}{N}, \quad i = 1,\ldots,N,
\end{align}
where $\z = ({\z}_1^{\top}, \ldots, {\z}_N^{\top})^{\top}$ is partitioned conformably according to ${\x}$, and $I_{\mathcal{Z}}({\z})$ is the indicator function of the convex set $\mathcal{Z}$, i.e.\ $I_{\mathcal{Z}}({\z}) = 0$ for ${\z} \in \mathcal{Z} = \{{\z}| \sum_{i=1}^{N}{\z}_i = 0\}$ and $I_{\mathcal{Z}}({\z}) = \infty$ otherwise. The augmented Lagrangian function is
\begin{align}
\mathcal{L}_{\rho} &= \sum_{i=1}^{N} f_i({\x}_i) + I_{\mathcal{Z}}({\z}) - \sum_{i=1}^{N}{\blambda}_{i}^{\top}({\A}_i{\x}_i + {\z}_i  - \frac{\mathbf{c}}{N}) \nonumber \\
& + \frac{\rho}{2}\sum_{i=1}^{N} \Vert {\A}_i{\x}_i + {\z}_i  - \frac{\mathbf{c}}{N} \Vert_2^2,
\end{align}
where we have two groups of variables, $\{{\x}_1, \ldots, {\x}_N\}$ and $\{{\z}_1, \ldots, {\z}_N\}$. Hence, we can apply the two-block ADMM to update these two groups of variables iteratively, i.e,  we can first update group $\{{\x}_i\}$ and then update group $\{{\z}_i\}$. In each group, ${\x}_i$ and ${\z}_i$ can be updated concurrently in parallel at each iteration. In particular, the update rules for ${\x}_i$ and ${\z}_i$ are
\begin{equation}
\left\{\begin{array}{ll}
{\x}_i^{k+1} = \argmin_{{\x}_i}\mathcal{L}_{\rho} ({\x}_i,{\z}_i^{k},{\blambda}_{i}^{k}),&\\
{\z}_i^{k+1} = \argmin_{{\z}_i}\mathcal{L}_{\rho} ({\x}_1^{k+1},{\z}_i,{\blambda}_{i}^{k}),& \forall i = 1,\ldots,N,\\
{\blambda}_{i}^{k+1} = {\blambda}_{i}^{k} - {\rho}({\A}_i{\x}_i + {\z}_i  - \frac{\mathbf{c}}{N}).&
\end{array} \right.
\end{equation}
The variable splitting ADMM is illustrated in Algorithm \ref{alg:A4}. The relationship between this splitting scheme and the Jacobian splitting scheme has been outlined in the following work \cite{BMX13}.  Algorithm \ref{alg:A4} enjoys the convergence rates of the 2-block ADMM. However, the number of variables and constraints will increase substantially when $N$ is large, which will impact the efficiency and incur significant burden for the computation.

\begin{algorithm}[t]
\caption{Variable Splitting Multi-block ADMM}\label{alg:A4}
\begin{algorithmic}
\STATE Initialize: ${\x}^{0}$, ${\z}^{0}$, ${\blambda}^{0}$, $\rho>0$;

\FOR{$k=0,1,\ldots$}

\FOR{$i=1,\ldots,N$}

\STATE \COMMENT{${\x}_i$, ${\z}_i$ and ${\blambda}_{i}$ are updated \textbf{concurrently}.}

\STATE ${\x}_i^{k+1} = \argmin_{{\x}_i}\mathcal{L}_{\rho} ({\x}_i,{\z}_i^{k},{\blambda}_{i}^{k})$;

\STATE ${\z}_i^{k+1} = \argmin_{{\z}_i}\mathcal{L}_{\rho} ({\x}_1^{k+1},{\z}_i,{\blambda}_{i}^{k})$;

\STATE ${\blambda}_{i}^{k+1} = {\blambda}_{i}^{k} - {\rho}({\A}_i{\x}_i + {\z}_i  - \frac{\mathbf{c}}{N})$;

\ENDFOR

\ENDFOR

%\STATE Output ${\x}$, ${\blambda}$;
\end{algorithmic}
\end{algorithm}

\subsection{ADMM with Gaussian Back Substitution}
\label{sec:GSADMM}
Many efforts have been made to improve the convergence of the Guass-Seidel type multi-block ADMM \cite{BXM12,MTXMSZ14}. In this part, we describe the ADMM with Gaussian back substitution \cite{BXM12}, which asserts that if a new iterate is generated by correcting the output of Algorithm \ref{alg:A2} with a Gaussian back substitution procedure, then the sequence of iterates converges to a solution of problem (\ref{eqn:intro}). We first define vector $\mathbf{v} =({\x}_2^{\top}, \ldots, {\x}_N^{\top}, {\blambda}^{\top})^{\top}$, vector $\tilde{\mathbf{v}} =(\tilde{\x}_2^{\top}, \ldots, \tilde{\x}_N^{\top}, \tilde{\blambda}^{\top})^{\top}$, matrix $\mathbf{H} = \text{diag}(\rho {\A}_2^{\top}{\A}_2,\ldots,\rho {\A}_N^{\top}{\A}_N,\frac{1}{\rho}\mathbf{I}_{m})$ and $\mathbf{M}$ as
\begin{equation}
\mathbf{M} =
\left( \begin{array}{ccccc}
\rho {\A}_2^{\top}{\A}_2 & 0 & \ldots & \ldots & 0 \\
\rho {\A}_3^{\top}{\A}_2 & \rho {\A}_3^{\top}{\A}_3 & \ddots & &\vdots \\
\vdots & \vdots & \ddots & \ddots & \vdots\\
\rho {\A}_N^{\top}{\A}_2 & \rho {\A}_N^{\top}{\A}_3 & \ldots & \rho {\A}_N^{\top}{\A}_N & 0\\
0&0&\ldots&0&\frac{1}{\rho}\mathbf{I}_{m}
\end{array} \right).
\end{equation}

Each iteration of the ADMM with Gaussian back substitution consists of two procedures: a prediction procedure and a correction procedure. The $\tilde{\mathbf{v}}$ is generated by the Algorithm \ref{alg:A2}. In particular, $\tilde{\x}_i$ is updated sequentially as
\begin{equation}
\tilde{\x}_i^{k} = \argmin_{\tilde{\x}_i}\mathcal{L}_{\rho}(\{\tilde{\x}_j^{k}\}_{j<i},{\x}_i,\{{\x}_j^{k}\}_{j>i},{\blambda}^k),
\end{equation}
where the prediction procedure is performed in a forward manner, i.e. from the first to the last block and to the Lagrangian multiplier. Note that the newly generated $\tilde{\x}_i$ are used in the update of the next block in accordance with the Gauss-Seidel update fashion. After the update of the Lagrangian multiplier, the correction procedure is performed update $\mathbf{v}$ as
\begin{equation}
\mathbf{H}^{-1}\mathbf{M}^{\top}(\mathbf{v}^{k+1}-\mathbf{v}^{k}) = \alpha(\tilde{\mathbf{v}}^{k}-\mathbf{v}^{k}),
\end{equation}
where $\mathbf{H}^{-1}\mathbf{M}^{\top}$ is a upper-triangular block matrix according to the definition of $\mathbf{H}$ and $\mathbf{M}$. This implies that the update of correction procedure is in a backward fashion, i.e, first update the Lagrangian multiplier, and then update ${\x}_i$ from the last block to the first block sequentially. Note that an additional assumption that ${\A}_i^{\top}{\A}_i(i = 1,2, \ldots, N)$ are nonsingular are made here. ${\x}_1$ serves as an intermediate variable and is unchanged during the correction procedure. The algorithm is illustrated in Algorithm \ref{alg:A5}.

The global convergence of the ADMM with Gaussian back substitution is proved in \cite{BXM12}, and the convergence rate and iteration complexity are addressed in \cite{BMX12}.

\begin{algorithm}[t]
\caption{ADMM with Gaussian Back Substitution}\label{alg:A5}
\begin{algorithmic}
\STATE Initialize: ${\x}^{0}$, $\tilde{\x}^{0}$, ${\blambda}^{0}$, $\tilde{\blambda}^{0}$, $\rho>0$, $\alpha \in (0,1)$;

\FOR{$k=0,1,\ldots$}

\FOR{$i=1,\ldots,N$}

\STATE \COMMENT{${\x}_i$ is updated \textbf{sequentially}.}

\STATE $\tilde{\x}_i^{k} = \argmin_{\tilde{\x}_i}\mathcal{L}_{\rho}(\{\tilde{\x}_j^{k}\}_{j<i},{\x}_i,\{{\x}_j^{k}\}_{j>i},{\blambda}^k)$;

\ENDFOR

\STATE $\tilde{\blambda}^{k+1} = {\blambda}^{k} - {\rho}(\sum_{i=1}^{N} {\A}_i\tilde{\x}_i^{k+1} - \mathbf{c})$;

\STATE

\STATE \COMMENT{Gaussian back substitution correction step}

\STATE $\mathbf{H}^{-1}\mathbf{M}^{\top}(\mathbf{v}^{k+1}-\mathbf{v}^{k}) = \alpha(\tilde{\mathbf{v}}^{k}-\mathbf{v}^{k})$;

\STATE ${\x}_1^{k+1} = \tilde{\x}_1^{k}$;
\ENDFOR

%\STATE Output ${\x}$, ${\blambda}$;
\end{algorithmic}
\end{algorithm}

\subsection{Proximal Jacobian ADMM}
\label{sec:PJADMM}
The other type of modification on the ADMM for the multi-block setting is based on the Jacobian iteration scheme \cite{BLX13,WMZW14,HAZ14,BMX13}. Since the Guass-Seidel update is performed sequentially and is not amenable for parallelization,  Jacobian type iteration is preferred for distributed and parallel optimization.  In this part we describe the proximal Jacobian ADMM \cite{WMZW14}, in which a proximal term \cite{NS13} is added in the update compare with that of Algorithm \ref{alg:A3} to improve convergence. In particular, the update of ${\x}_i$ is
\begin{equation}
{\x}_i^{k+1} \! = \! \argmin_{{\x}_i}\mathcal{L}_{\rho}({\x}_i,\{{\x}_j^{k}\}_{j \neq i},{\blambda}^k) \! + \! \frac{1}{2}\Vert {\x}_i \! - \! {\x}_i^{k}\Vert_{\mathbf{P}_i}^{2},
\end{equation}
where $\Vert {\x}_i \Vert_{\mathbf{P}_i}^{2} = {\x}_i^{\top}\mathbf{P}_i{\x}_i$ for some symmetric and positive semi-definite matrix $\mathbf{P}_i \succeq 0$. The involvement of the proximal term can make the subproblem of ${\x}_i$ strictly or strongly convex and thus make the problem more stable. Moreover, multiple choice of $\mathbf{P}_i$ can make the subproblems easier to solve. The update of Lagrangian multiplier is
\begin{equation}
{\blambda}^{k+1} = {\blambda}^{k} - {\gamma}{\rho}(\sum_{i=1}^{N} {\A}_i{\x}_i^{k+1} - \mathbf{c}),
\end{equation}
where $\gamma >0$ is the damping parameter and the algorithm is illustrate in Algorithm \ref{alg:A6}.

The global convergence of the proximal Jacobian ADMM which is proved in \cite{WMZW14}. Moreover, it enjoys a convergence rate of $o(1/k)$ under conditions on $\mathbf{P}_i$ and $\gamma$.

\begin{algorithm}[t]
\caption{Proximal Jacobian ADMM}\label{alg:A6}
\begin{algorithmic}
\STATE Initialize: ${\x}^{0}$, ${\blambda}^{0}$, $\rho>0$, $\gamma >0$;

\FOR{$k=0,1,\ldots$}

\FOR{$i=1,\ldots,N$}

\STATE \COMMENT{${\x}_i$ is updated \textbf{concurrently}.}

\STATE ${\x}_i^{k+1} \! = \! \argmin_{{\x}_i}\mathcal{L}_{\rho}({\x}_i,\{{\x}_j^{k}\}_{j \neq i},{\blambda}^k) \! + \! \frac{1}{2}\Vert {\x}_i \! - \! {\x}_i^{k}\Vert_{\mathbf{P}_i}^{2}$;

\ENDFOR

\STATE ${\blambda}^{k+1} = {\blambda}^{k} - {\gamma}{\rho}(\sum_{i=1}^{N} {\A}_i{\x}_i^{k+1} - \mathbf{c})$;

\ENDFOR
%\STATE Output ${\x}$, ${\blambda}$;
\end{algorithmic}
\end{algorithm}

\subsection{Implementations}
The recent development in high performance computing (HPC) and cloud computing paradigm provides a flexible and efficient solution for deploying the large-scale optimization algorithms. In this part, we describe possible implementation approaches of those distributed and parallel algorithms on current mainstream large scale computing facilities.

One possible implementation utilizes available computing incentive techniques and tools like MPI, OpenMP, and OpenCL. The MPI is a language-independent protocol used for inter-process communications on distributed memory computing platform, and is widely used for high-performance parallel computing today. The (multi-block) ADMM using MPI has been implemented in \cite{BPCPE10} and \cite{ZMW13}. Besides, the OpenMP, which is a shared memory multiprocessing parallel computing paradigm, and the OpenCL, which is a heterogenous distributed-shared memory parallel computing paradigm that incorporate CPUs and GPUs, also promise to implement distributed and parallel optimization algorithms on HPC. It is expected that supercomputers will reach one exaflops ($10^{18}$ FLOPS) and even zettaflops ($10^{21}$ FLOPS) in the near feature, which will largely enhance the computing capacity and significantly expedite program execution.

Another possible approach exploits the ease-of-use cloud computing engine like Hadoop MapReduce and Apache Spark. The amount of cloud infrastructure available for Hadoop MapReduce makes it convenient to use for large problems, though it is awkward to express ADMM in MapReduce since it is not designed for iterative tasks. Apache Spark's in-memory computing feature enables it to run iterative optimizations much faster than Hadoop, and is now prevalent for large-scale machine learning and optimization task on clusters\cite{MMMSI10}. This implementation approach is much simpler than previous computing incentive techniques and tools and promise to implementation of the large-scale distributed and parallel computation algorithms based on ADMM. The advances in the cloud/cluster computing engine provides a simple method to implement the large-scale data processing, and recently Google, Baidu and Alibaba are also developing and deploying massive cluster computing engines to perform the large-scale distributed and parallel computation.

\begin{figure}[t]
    \centering
    \includegraphics[width=0.43\textwidth]{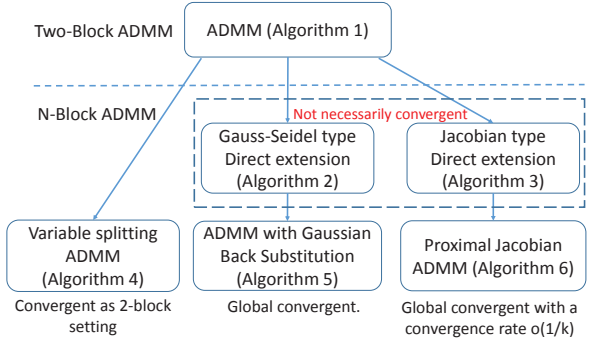}
    \caption{An illustration of the relationships between Algorithms 1-6.}
    \label{fig:alg}
\end{figure}

Now we have finished the review of distributed and parallel optimization methods based on ADMM, and we summarize the relationships between Algorithms $1-6$ in Fig. \ref{fig:alg}.

\section{Smart Grid Applications}
\label{sec:application}
In this section, we review several applications of distributed and parallel optimization in smart grid communication networks for distributed robust state estimation \cite{VG13, LMQVZ14}, network energy management \cite{MEJS13} and security constrained optimal power flow problem \cite{LAWZ13,SMERS14} based on ADMM.

\subsection{Distributed Robust Power System State Estimation}
In this subsection, we consider the robust state estimation in power system \cite{VG13, LMQVZ14}. State estimation, which estimates the power system operating states based on a real-time electric network model, is a key function of the energy management system. Assume an interconnected power system consisting of $N$ control areas. Each control area has its own control center which estimate the sub-system states, and the whole system operating states can be obtained by inter-area communications between control centers. Additionally, the state estimation scheme should be able to detect false data injection into the power system. In \cite{VG13}, the distributed robust power system state estimation can be formulated as
\begin{align}
\label{eqn:RobustSE}
\min_{\{{\x}_i \in \mathcal{X}_i\},\{{\so}_i\}}\quad & \sum_{i=1}^{N}( \frac{1}{2} \Vert {\m}_i - {\J}_i{\x}_i - {\so}_i\Vert_2^{2} + \beta \Vert {\so}_i \Vert_1) ,\nonumber \\
\text{s.t.}\quad & {\x}_i[j] = {\x}_j[i], \quad \forall j \in \mathcal{N}_{i}, \quad \forall i,
\end{align}
where ${\m}_i$ is the state measurement aggregated at each control center and ${\x}_i$ is the sub-system state at the $i^{th}$ control area, respectively. ${\J}_i$ is the Jacobian matrix and ${\so}_i$ is the injected false data. Note that the false data injection vector is sparse and thus a $l_1$ norm sparse regularization term is employed in the problem (\ref{eqn:RobustSE}) to enforce the sparsity of ${\so}_i$. The constraints of problem (\ref{eqn:RobustSE}) require the neighboring areas to consent on their shared states, where $\mathcal{N}_{i}$ denotes the set of neighbors of area $i$. The optimization problem (\ref{eqn:RobustSE}) is solved in a distributed fashion by ADMM, and the optimal solution recovers the system states as well as detects the false date injection vector.

\subsection{Dynamic Network Energy Management}
In this subsection, we consider the dynamic network energy management \cite{MEJS13}. Assume a network of devices, such as generators, loads, and storage devices, connected by AC and DC lines. The goal is to jointly minimize a network objective subject to local constraints on the devices and lines. Let $\mathcal{D}$ and $\mathcal{N}$ denote the set of devices and nets in the power system, respectively. The dynamic network energy management can be formulated as an optimization problem as

\begin{align}
\label{eqn:DM}
\min &\quad \sum_{d \in \mathcal{D}} f_d(p_d, \theta_d) + \sum_{n \in \mathcal{N}}(g_n(z_n) + h_n({\xi}_n)), \nonumber \\
\text{s.t.} &\quad p = z, \quad \theta = \xi,
\end{align}

where $p_d$ and $\theta_d$ are power schedules and phase schedules associated with device $d$, respectively. The function $f_d(p_d, \theta_d)$ represents the cost (or revenue, if negative) to device d for operating according to power and phase schedule. The function $g_n(z_n)$ is the indicator function on the set $\{z_n|\bar{z_n} = 0\}$, where $\bar{z_n}$ denotes the average net power imbalance. The function $h_n({\xi}_n)$ is the indicator function on the set $\{{\xi}_n|\tilde{{\xi}_n} = 0\}$, where $\tilde{{\xi}_n}$ denotes the phase residual of the power system. These two functions enforce the power balance and phase consistency holds across all nets. The auxiliary variables $\{z_n\}$ and $\{{\xi}_n\}$ are introduced to facilitate the parallelization of the problem (\ref{eqn:DM}). The optimization problem (\ref{eqn:DM}) can be solved in a fully decentralized manner based on ADMM by message passing between devices in the system, and the optimal value, optimal power and phase schedules as well as locational marginal prices can be obtained.

\subsection{Security Constrained Optimal Power Flow}
In this subsection, we consider the distributed and parallel approach for security constrained optimal power flow problem (SCOPF) \cite{LAWZ13, SMERS14}. The SCOPF is an extension of the conventional optimal power flow (OPF) problem, whose objective is to determine a generation schedule that minimizes the system operating cost while satisfying the system operation constraints such as hourly load demand, fuel limitations, environmental constraints and network security requirements. In \cite{LAWZ13}, the general form of SCOPF can be formulated as follows
\begin{align}
\label{SCOPF}\min_{\mathbf{x}^0,\ldots,\mathbf{x}^C;\mathbf{u}^0,\ldots,\mathbf{u}^C} & \quad f^0(\mathbf{x}^0,\mathbf{u}^0) + \sum_{c=1}^{C} I_c(\mathbf{x}^c,\mathbf{u}^c)\\
\text{s.t.} & \label{pos3} \quad \vert \mathbf{u}^0 - \mathbf{u}^c \vert \le \mathbf{\Delta}_c, \quad  c = 1,\ldots,C,
\end{align}
where $f^0$ is the objective function, through which (\ref{SCOPF}) aims to maximize the total social welfare or equivalently minimize the offer-based energy and production cost. $\mathbf{x}^c$ is the vector of state variables, which includes magnitude and voltage angle at all buses, and $\mathbf{u}^c$ is the vector of control variables, which can be generators' real powers or terminal voltages. The superscript $c = 0$ corresponds to the pre-contingency configuration, and $c = 1,\ldots,C$ correspond to different post-contingency configurations. The function $I_c(\mathbf{x}^c,\mathbf{u}^c)$ is the indicator function on the set $\{ (\mathbf{x}^c,\mathbf{u}^c) | \mathbf{g}^c(\mathbf{x}^c,\mathbf{u}^c) = 0, \mathbf{h}^c(\mathbf{x}^c,\mathbf{u}^c) \le 0\}$. The equality constraints $\mathbf{g}^c, c = 0,\ldots,C$ represent the system nodal power flow balance over the entire grid, and inequality constraints $\mathbf{h}^c, c = 0,\ldots,C$ represent the physical limits on the equipment. $\mathbf{\Delta}_c$ is the maximal allowed adjustment between the normal and contingency states for contingency $c$.  The SCOPF problem  is solved in a distributed fashion by ADMM, and the optimal solution finds the optimal generation schedule subjects to the contingency constraints.

\section{Summary}
\label{sec:conclusion}
In this paper, we have reviewed several distributed and parallel optimization method based on the ADMM for large scale optimization problem. We have introduced the background of ADMM and described several direct extensions and sophisticated modifications of ADMM from $2$-block to $N$-block settings. We have explained the iterative schemes and convergence properties for each extension/modification. We have illustrated the implantations on large-scale computing facilities, and enumerated several applications of $N$-block ADMM in smart grid communication networks.

\balance
\bibliographystyle{IEEEtran}
\bibliography{JOC}

\end{document}